\def\aa{{A\&A}}

\def\aj{{AJ}}

\def\apj{{ApJ}}

\def\mnras{{MNRAS}}
\def\nat{{Nature}}

\def\lsim{\mathrel{\rlap{\lower 4pt \hbox{\hskip 1pt $\sim$}}\raise 1pt
\hbox {$<$}}} 
\def\gsim{\mathrel{\rlap{\lower 4pt \hbox{\hskip 1pt $\sim$}}\raise 1pt
\hbox {$>$}}}
\newcommand{\ms}{$M_\odot$}
\newcommand{\Nifs}{$^{56}$Ni}

\documentclass[cup5b]{caps}
\usepackage{graphicx}
\usepackage{amssymb}
\usepackage{ociwsymp4e}  

\HeadText{K. Nomoto et al.}  

\begin{document}

\pagenumbering{arabic}

\author[]{K. NOMOTO$^{1}$, K. MAEDA$^{1}$, 
H. UMEDA$^{1}$, 
N. TOMINAGA$^{1}$, 
T. OHKUBO$^{1}$, 
J. DENG$^{1}$, 
P. A. MAZZALI$^{1,2}$
\\
(1) Department of Astronomy \& Research Center for the Early Universe, University of Tokyo, Tokyo, Japan\\
(2) Osservatorio Astronomico, Via Tiepolo, 11, 34131 Trieste, Italy
}

%
%

\chapter{Nucleosynthesis in Black-Hole-Forming Supernovae and 
Abundance Patterns of Extremely Metal-Poor Stars}

\begin{abstract}

Stars more massive than $\sim$ 20 - 25 \ms\ form a black hole at the
end of their evolution.  Stars with non-rotating black holes are
likely to collapse "quietly" ejecting a small amount of heavy elements
(Faint supernovae).  In contrast, stars with rotating black holes are
likely to give rise to very energetic supernovae (Hypernovae).  We
present distinct nucleosynthesis features of these two types of
"black-hole-forming" supernovae.  Nucleosynthesis in Hypernovae is
characterized by larger abundance ratios (Zn,Co,V,Ti)/Fe and smaller
(Mn,Cr)/Fe than normal supernovae, which can explain the observed
trend of these ratios in extremely metal-poor stars.  Nucleosynthesis
in Faint supernovae is characterized by a large amount of fall-back.
We show that the abundance pattern of the recently discovered most
Fe-poor star, HE0107-5240, and other extremely metal-poor carbon-rich
stars are in good accord with those of black-hole-forming supernovae,
but not pair-instability supernovae.  This suggests that
black-hole-forming supernovae made important contributions to the
early Galactic (and cosmic) chemical evolution.  Finally we discuss
the nature of First (Pop III) Stars.

\end{abstract}

\section{Introduction}

Among the important developments in recent studies of core-collapse
supernovae are the discoveries of two distinct types of supernovae
(SNe): 1) very energetic SNe (Hypernovae), whose kinetic energy (KE)
exceeds $10^{52}$\,erg, about 10 times the KE of normal core-collapse
SNe (hereafter $E_{51} = E/10^{51}$\,erg), and 2) very faint and low
energy SNe ($E_{51} \lsim$ 0.5; Faint supernovae).  These two types of
supernovae are likely to be "black-hole-forming" supernovae with
rotating or non-rotating black holes.  We compare their
nucleosynthesis yields with the abundances of extremely metal-poor
(EMP) stars to identify the Pop III (or first) supernovae.  We show
that the EMP stars, especially the C-rich class, are likely to be
enriched by black-hole-forming supernovae.

\begin{figure}
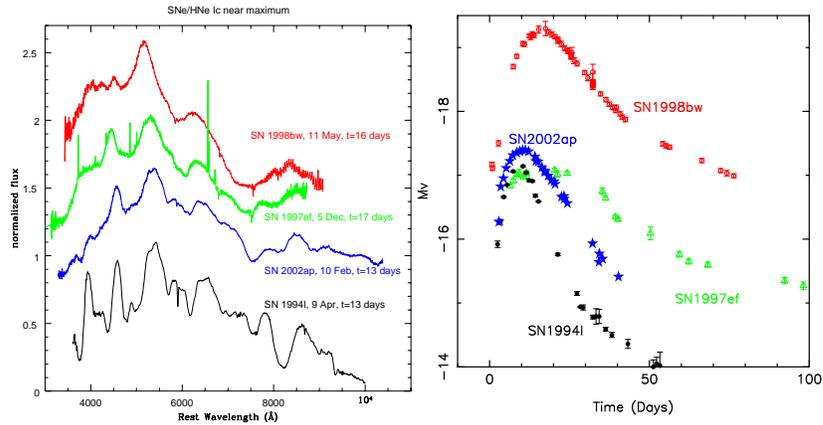

	\begin{center}
		\begin{minipage}[t]{0.46\textwidth}
			\includegraphics[width=1.0\textwidth]{f1s02ap.epsi}
		\end{minipage}
		\begin{minipage}[t]{0.46\textwidth}
			\includegraphics[width=1.0\textwidth]{f2s02ap.epsi}
		\end{minipage}
	\end{center}
	\caption{Left: The near-maximum spectra of Type Ic SNe and
 hypernovae: SNe 1998bw, 1997ef, 2002ap, and 1994I. Right: 
The observed $V$-band light curves of SNe 1998bw ({\em open
circles}), 1997ef ({\em open triangles}), 2002ap ({\em stars}), and
1994I ({\em filled circles}) (Mazzali et al. 2002). 
\label{fig1}}
\end{figure}

\begin{figure}
	\begin{center}
		\begin{minipage}[t]{0.65\textwidth}
			\includegraphics[width=1.\textwidth]{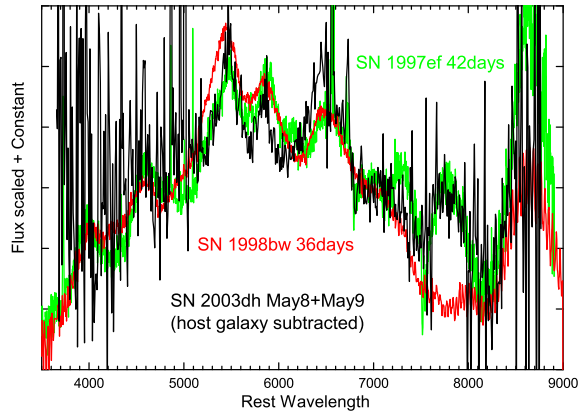}
		\end{minipage}
	\end{center}
	\caption{Spectrum of SN 2003dh/GRB 030329 
compared with SNe 1998bw and 1997ef (Kawabata et al. 2003).
\label{fig11}}
\end{figure}

\section{Hypernova Branch and Faint Supernova Branch}

Type Ic Hypernova SN 1998bw was probably linked to GRB 980425
(Galama et al. 1998), thus establishing for the first time a connection
between gamma-ray bursts (GRBs) and core-collapse SNe.  However,
SN~1998bw was exceptional for a SN~Ic: it was as luminous at peak as a
SN~Ia, indicating that it synthesized $\sim 0.5$ \ms\ of \Nifs, and
its KE was estimated as $E \sim 3 \times 10^{52}$ erg
(Iwamoto et al. 1998; Woosley, Eastman, \& Schmidt 1999). 

Subsequently, other ``hypernovae" have been recognized, such as
SN~1997ef (Iwamoto et al. 2000; Mazzali, Iwamoto, \& Nomoto 2000),
SN~1999as (Knop et al. 1999; Hatano et al. 2001), and SN~2002ap
(Mazzali et al. 2002).  Although these SNe Ic did not appear to be
associated with GRBs, most recent ``hypernova'' SN 2003dh is clearly
associated with GRB 030329 (Stanek et al. 2003; Kawabata et al. 2003).
Figure~\ref{fig1} shows the near-maximum spectra and the absolute
$V$-light curves of hypernovae.  Figure~\ref{fig11} shows that the
spectrum of SN 2003dh/GRB 030329 is in good agreement with SNe 1998bw
and 1997ef, especially 97ef (Kawabata et al. 2003).  These hypernovae
span a wide range of properties, although they all appear to be highly
energetic compared to normal core-collapse SNe.

\begin{figure}
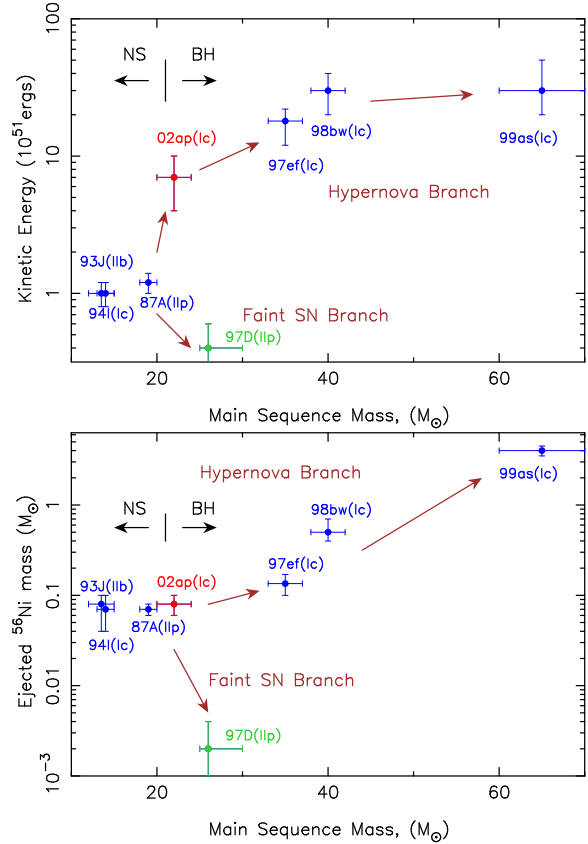

\begin{center}
  \begin{minipage}[t]{0.65\textwidth}
		\includegraphics[width=0.92\textwidth]{e-m_c.epsi}
  \end{minipage}
  \begin{minipage}[t]{0.65\textwidth}
		\includegraphics[width=0.92\textwidth]{ni-m_c.epsi}
  \end{minipage}
\end{center}
\caption{
The explosion energy and the ejected $^{56}$Ni mass as a function of
the main sequence mass of the progenitors for several
supernovae/hypernovae (Nomoto et al. 2003ab).
\label{fig2}}
\end{figure}

Figure~\ref{fig2} shows $E$ and the mass of $^{56}$Ni ejected
$M(^{56}$Ni) as a function of the main-sequence mass $M_{\rm ms}$ of
the progenitor star obtained from fitting the optical light curves and
spectra.  These mass estimates place hypernovae at the high-mass end
of SN progenitors.

In contrast, SNe II 1997D and 1999br were very faint SNe with very low
$KE$ (Turatto et al. 1998; Hamuy 2003; Zampieri et al. 2003).  In
Figure~\ref{fig2}, therefore, we propose that SNe from stars with
$M_{\rm ms} \gsim$ 20-25 \ms\ have different $E$ and $M(^{56}$Ni),
with a bright, energetic ``hypernova branch'' at one extreme and a
faint, low-energy SN branch at the other (Nomoto et al. 2003ab).  For
the faint SNe, the explosion energy was so small that most \Nifs\ fell
back onto the compact remnant.  Thus the faint SN branch may become a
``failed'' SN branch at larger $M_{\rm ms}$.  Between the two
branches, there may be a variety of SNe (Hamuy 2003).

This trend might be interpreted as follows.  Stars with $M_{\rm ms}
\lsim$ 20-25 \ms\ form a neutron star, producing $\sim$ 0.08 $\pm$
0.03 \ms\ \Nifs\ as in SNe 1993J, 1994I, and 1987A (SN 1987A may be a
borderline case between the neutron star and black hole formation).
Stars with $M_{\rm ms} \gsim$ 20-25 \ms\ form a black hole; whether
they become hypernovae or faint SNe may depend on the angular momentum
in the collapsing core, which in turn depends on the stellar winds,
metallicity, magnetic fields, and binarity.  Hypernovae might have
rapidly rotating cores owing possibly to the spiraling-in of a
companion star in a binary system.

\section{Nucleosynthesis in Hypernova Explosions}

In core-collapse supernovae/hypernovae, stellar material undergoes
shock heating and subsequent explosive nucleosynthesis. Iron-peak
elements are produced in two distinct regions, which are characterized
by the peak temperature, $T_{\rm peak}$, of the shocked material.  For
$T_{\rm peak} > 5\times 10^9$K, material undergoes complete Si burning
whose products include Co, Zn, V, and some Cr after radioactive
decays.  For $4\times 10^9$K $<T_{\rm peak} < 5\times 10^9$K,
incomplete Si burning takes place and its after decay products include
Cr and Mn (Hashimoto, Nomoto, \& Shigeyama 1989; Thielemann, Nomoto,
\& Hashimoto 1996).

\subsection {Supernovae vs. Hypernovae}

\begin{figure}[t]
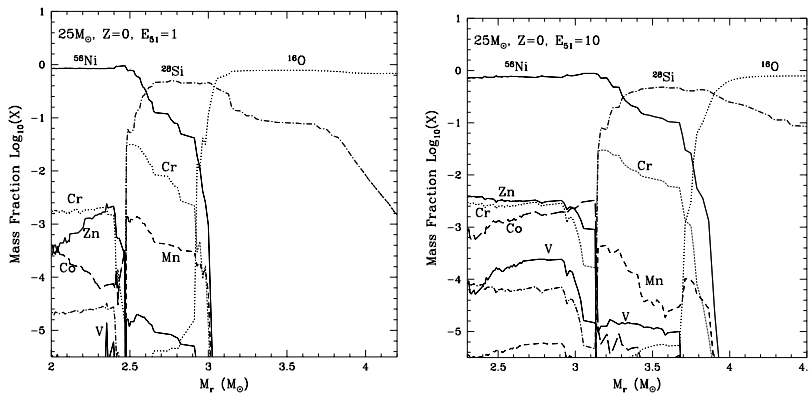

 \begin{center}
\begin{minipage}[t]{0.47\textwidth}
		\includegraphics[width=0.95\textwidth]{fig3a.epsi}
\end{minipage}
\begin{minipage}[t]{0.47\textwidth}
		\includegraphics[width=0.95\textwidth]{fig3b.epsi}
\end{minipage}
 \end{center}
 \caption[]{Abundance distribution plotted against the enclosed mass
$M_r$ after the explosion of Pop III 25 \ms\ stars with $E_{51} = 1$
(left) and $E_{51} = 10$ (right) (Umeda \& Nomoto 2002a). }
\label{fig3}
\end{figure}

The right panel of Figure~\ref{fig3} shows the composition in the
ejecta of a 25 \ms\ hypernova model ($E_{51} = 10$).  The
nucleosynthesis in a normal 25 \ms\ SN model ($E_{51} = 1$) is also
shown for comparison in the left panel of Figure~\ref{fig3} (Umeda \&
Nomoto 2002a).

We note the following characteristics of nucleosynthesis with very
large explosion energies (Nakamura et al. 2001b; Nomoto et al. 2001ab;
Ohkubo, Umeda, \& Nomoto 2003):

(1) Both complete and incomplete Si-burning regions shift outward in
mass compared with normal supernovae, so that the mass ratio between
the complete and incomplete Si-burning regions becomes larger.  As a
result, higher energy explosions tend to produce larger [(Zn, Co,
V)/Fe] and smaller [(Mn, Cr)/Fe], which can explain the trend observed
in very metal-poor stars (Umeda \&  Nomoto 2002b, 2003b).

(2) In the complete Si-burning region of hypernovae, elements produced
by $\alpha$-rich freezeout are enhanced.  Hence, elements synthesized
through capturing of $\alpha$-particles, such as $^{44}$Ti, $^{48}$Cr,
and $^{64}$Ge (decaying into $^{44}$Ca, $^{48}$Ti, and $^{64}$Zn,
respectively) are more abundant.

(3) Oxygen burning takes place in more extended regions for the larger
KE.  Then more O, C, Al are burned to produce a larger amount of
burning products such as Si, S, and Ar.  Therefore, hypernova
nucleosynthesis is characterized by large abundance ratios of [Si,S/O],
which can explain the abundance feature of M82 (Umeda et al. 2002).

\subsection{Hypernovae and Zn, Co, Mn, Cr}

Hypernova nucleosynthesis may have made an important contribution to
Galactic chemical evolution.  In the early galactic epoch when the
galaxy was not yet chemically well-mixed, [Fe/H] may well be
determined by mostly a single SN event (Audouze \& Silk 1995). The
formation of metal-poor stars is supposed to be driven by a supernova
shock, so that [Fe/H] is determined by the ejected Fe mass and the
amount of circumstellar hydrogen swept-up by the shock wave (Ryan,
Norris, \& Beers 1996).  Then, hypernovae with larger $E$ are likely
to induce the formation of stars with smaller [Fe/H], because the mass
of interstellar hydrogen swept up by a hypernova is roughly
proportional to $E$ (Ryan et al. 1996; Shigeyama \& Tsujimoto 1998)
and the ratio of the ejected iron mass to $E$ is smaller for
hypernovae than for normal supernovae.

In the observed abundances of halo stars, there are significant
differences between the abundance patterns in the iron-peak elements
below and above [Fe/H]$ \sim -2.5$ - $-3$.

(1) For [Fe/H]$\lsim -2.5$, the mean values of [Cr/Fe] and [Mn/Fe]
decrease toward smaller metallicity, while [Co/Fe] increases
(McWilliam et al. 1995; Ryan et al. 1996).

(2) [Zn/Fe]$ \sim 0$ for [Fe/H] $\simeq -3$ to $0$ (Sneden, Gratton,
\& Crocker 1991), while at [Fe/H] $< -3.3$, [Zn/Fe] increases toward
smaller metallicity (Primas et al. 2000; Blake et al. 2001).

The larger [(Zn, Co)/Fe] and smaller [(Mn, Cr)/Fe] in the supernova
ejecta can be realized if the mass ratio between the complete Si
burning region and the incomplete Si burning region is larger, or
equivalently if deep material from the complete Si-burning region is
ejected by mixing or aspherical effects.  This can be realized if (1)
the mass cut between the ejecta and the compact remnant is located at
smaller $M_r$ (Nakamura et al. 1999), (2) $E$ is larger to move the
outer edge of the complete Si burning region to larger $M_r$ (Nakamura
et al. 2001a), or (3) asphericity in the explosion is larger.

Among these possibilities, a large explosion energy $E$ enhances
$\alpha$-rich freezeout, which results in an increase of the local
mass fractions of Zn and Co, while Cr and Mn are not enhanced (Umeda
\& Nomoto 2002ab; Ohkubo et al. 2003).  Models with $E_{51} = 1 $ do
not produce sufficiently large [Zn/Fe].  To be compatible with the
observations of [Zn/Fe] $\sim 0.5$, the explosion energy must be much
larger, i.e., $E_{51} \gsim 20$ for $M \gsim 20 M_\odot$, i.e.,
hypernova-like explosions of massive stars ($M \gsim 25 M_\odot$) with
$E_{51} > 10$ are responsible for the production of Zn.

\begin{figure}[t]
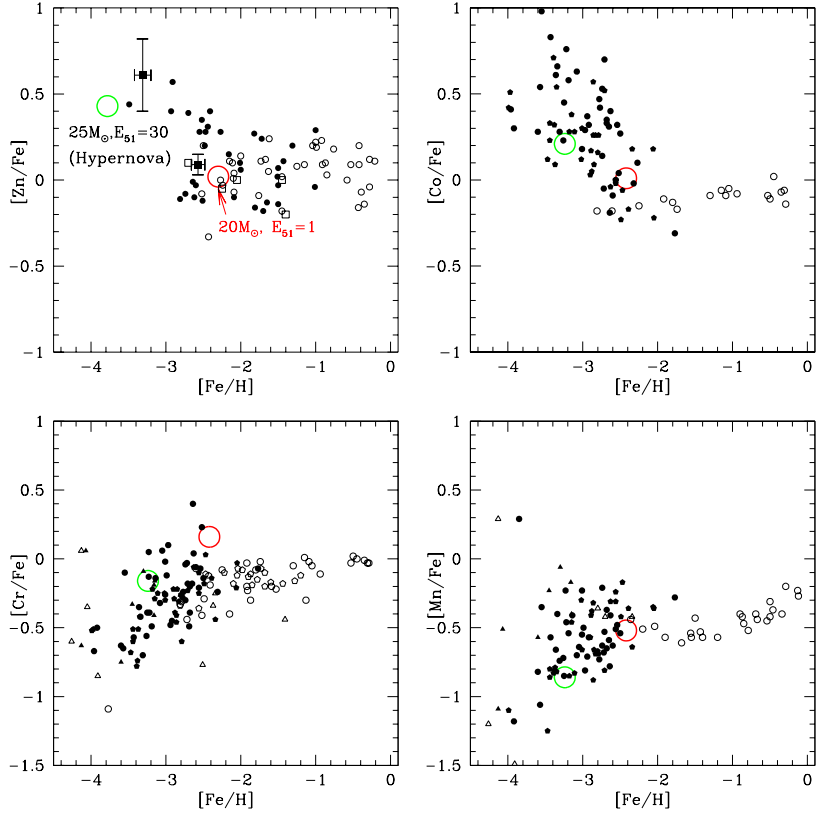

 \begin{center}
\begin{minipage}[t]{0.47\textwidth}
		\includegraphics[width=0.95\textwidth]{znfe2C.epsi}
\end{minipage}
\begin{minipage}[t]{0.47\textwidth}
		\includegraphics[width=0.95\textwidth]{cofe2C.epsi}
\end{minipage}
 \end{center}
 \begin{center}
\begin{minipage}[t]{0.47\textwidth}
		\includegraphics[width=0.95\textwidth]{crfe5001_4996C.epsi}
\end{minipage}
\begin{minipage}[t]{0.47\textwidth}
		\includegraphics[width=0.95\textwidth]{mnfe5001_4996C.epsi}
\end{minipage}
 \end{center}
 \caption[]{Observed abundance ratios of [Zn, Co, Cr, Mn/Fe] vs [Fe/H]
compared with (15M, $E_{51}=1$) and (25M, $E_{51}$=30) models (Umeda
\& Nomoto 2003b).  
}
\label{figznfe}
\end{figure}

In the hypernova models, the overproduction of Ni, as found in the
simple ``deep'' mass-cut model, can be avoided (Ohkubo et al. 2003).
Therefore, if hypernovae made significant contributions to the early
Galactic chemical evolution, it could explain the large Zn and Co
abundances and the small Mn and Cr abundances observed in very
metal-poor stars (Fig.~\ref{figznfe}: Umeda \& Nomoto 2003b).

\section{Extremely Metal-Poor (EMP) Stars and Faint Supernovae}

Recently the most Fe deficient and C-rich low mass star, HE0107-5240,
was discovered (Christlieb et al. 2002).  This star has [Fe/H] $= -
5.3$ but its mass is as low as 0.8 \ms.  This would challenge the
recent theoretical arguments that the formation of low mass stars,
which should survive until today, is suppressed below [Fe/H] $= -4$
(Schneider et al. 2002).

The important clue to this problem is the observed abundance pattern
of this star.  This star is characterized by a very large ratios of
[C/Fe] = 4.0 and [N/Fe] = 2.3, while the abundances of elements
heavier than Mg are as low as Fe (Christlieb et al. 2002).
Interestingly, this is not the only extremely metal poor (EMP) stars
that have the large C/Fe and N/Fe ratios, but several other such stars
have been discovered (Ryan 2002).  Therefore the reasonable
explanation of the abundance pattern should explain other EMP stars as
well.  We show that the abundance pattern of C-rich EMP stars can be
reasonably explained by the nucleosynthesis of 20 - 130 \ms\
supernovae with various explosion energies and the degree of mixing
and fallback of the ejecta.

\begin{figure}[t]
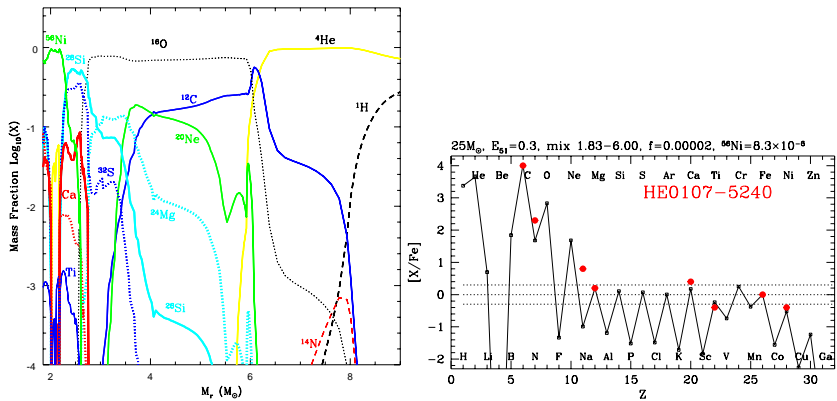

  \begin{center}
  \begin{minipage}[t]{0.45\textwidth}
		\includegraphics[width=1.\textwidth]{25z0nat.epsi}
  \end{minipage}
  \begin{minipage}[t]{0.49\textwidth}
		\includegraphics[width=1.\textwidth]{he0107.epsi}
  \end{minipage}
 \end{center}
\caption{
(left) The post-explosion abundance distributions for the 25 $M_\odot$
model with the explosion energy $E_{51} =$ 0.3 (Umeda \& Nomoto
2003a).  (right) Elemental abundances of the C-rich most Fe deficient
star HE0107-5240 (filled circles), compared with a theoretical
supernova yield (Umeda \& Nomoto 2003a).
\label{fig4}}
\end{figure}

\subsection{The Most Fe-Poor Star HE0107-5240}

We consider a model that C-rich EMP stars are produced in the ejecta
of (almost) metal-free supernova mixed with extremely metal-poor
interstellar matter.  We use Pop III pre-supernova progenitor models,
simulate the supernova explosion and calculate detailed
nucleosynthesis (Umeda \& Nomoto 2003a).

In Figure~\ref{fig4} (right) we show that the elemental abundances of
one of our models are in good agreement with HE0107-5240, where the
progenitor mass is 25 \ms\ and the explosion energy $E_{51} =$ 0.3
(Umeda \& Nomoto 2003a).

In this model, explosive nucleosynthesis takes place behind the shock
wave that is generated at $M_r =$ 1.8 \ms\ and propagates outward. The
resultant abundance distribution is seen in Figure~\ref{fig4} (left),
where $M_r$ denotes the Lagrangian mass coordinate measured from the
center of the pre-supernova model (Umeda \& Nomoto 2003a).  The
processed material is assumed to mix uniformly in the region from $M_r
=$ 1.8 \ms\ and 6.0 \ms.  Such a large scale mixing was found to take
place in SN1987A and various explosion models (Hachisu et al. 1990;
Kifonidis et al. 2000). Almost all materials below $M_r =$ 6.0 \ms\
fall back to the central remnant and only a small fraction ($f = 2
\times$ 10$^{-5}$) is ejected from this region.  The ejected Fe mass
is 8 $\times$ 10$^{-6}$ \ms.

The CNO elements in the ejecta were produced by pre-collapse He shell
burning in the He-layer, which contains 0.2 \ms\ $^{12}$C.  Mixing of H
into the He shell-burning region produces 4 $\times$ 10$^{-4}$ \ms\
$^{14}$N.  On the other hand, only a small amount of heavier elements
(Mg, Ca, and Fe-peak elements) are ejected and their abundance ratios
are the average in the region of $M_r =$ 1.8 - 6.0 \ms. The sub-solar
ratios of [Ti/Fe] $= -0.4$ and [Ni/Fe] $= -0.4$ are the results of the
relatively small explosion energy ($E_{51} =$ 0.3).  With this "mixing
and fallback", the large C/Fe and C/Mg ratios observed in HE0107-5240
are well reproduced (Umeda \& Nomoto 2003a).

In this model, N/Fe appears to be underproduced. However, N can be
produced inside the EMP stars through the C-N cycle, and brought up to
the surface during the first dredge up stage while becoming a red-giant
star (Boothroyd \& Sackmann 1999).
 
\begin{figure}[t]
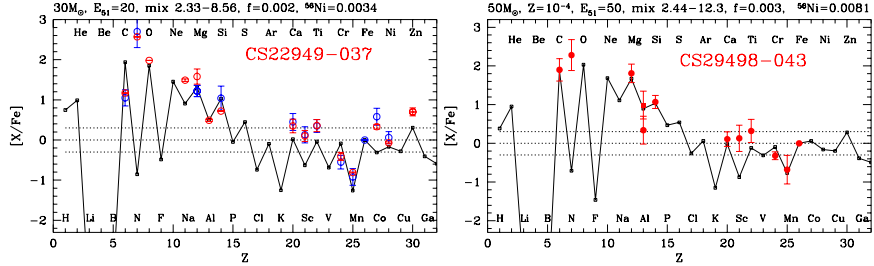

  \begin{center}
  \begin{minipage}[t]{0.49\textwidth}
		\includegraphics[width=1.\textwidth]{cs-037.epsi}
  \end{minipage}
  \begin{minipage}[t]{0.49\textwidth}
		\includegraphics[width=1.\textwidth]{cs-043.epsi}
  \end{minipage}  \end{center}
\caption{
(left) Elemental abundances of CS 22949-037 (open circles for Norris
et al. 2001, and solid squares for Depagne et al. 2002), compared with
a theoretical supernova yield (Umeda \& Nomoto 2003ab). (right)
Same as the left panel but for CS 29498-043 (Aoki et al. 2002).
\label{fig5}}
\end{figure}

\subsection{Carbon-rich EMP stars: CS 22949-037 and CS 29498-043} 

The "mixing and fallback" is commonly required to reproduce the
abundance pattern of typical EMP stars.  In Figure~\ref{fig5} (left)
we show a model, which is in good agreement with CS22949-037 (Umeda \&
Nomoto 2003a).  This star has [Fe/H] $= -4.0$ and also C, N-rich
(Norris2001, Ryan, \& Beers 2001; Depagne et al. 2002), though C/Fe
and N/Fe ratios are smaller than HE0107-5240.  The model is the
explosion of a 30 \ms\ star with $E_{51} =$ 20.  In this model, the
mixing region ($M_r =$ 2.33 - 8.56 \ms) is chosen to be smaller than
the entire He core ($M_r =$ 13.1 \ms) in order to reproduce relatively
large Mg/Fe and Si/Fe ratios.  

Similar degree of the mixing, but for a more massive progenitor, would
also reproduce the abundances of CS29498-043 (Aoki et al. 2002), which
shows similar abundance pattern (Fig.~\ref{fig5}: left).

We assume a larger fraction of ejection than HE0107-5240, 2\%, from
the mixed region for CS22949-037, because the C/Fe and N/Fe ratios
are smaller. The ejected Fe mass is 0.003 \ms. The larger explosion
energy model is favored for explaining the large Zn/Fe, Co/Fe and
Ti/Fe ratios (Umeda \& Nomoto 2002a).

Without mixing, elements produced in the deep explosive burning
regions, such as Zn, Co, and Ti, would be underproduced. Without
fallback the abundance ratios of heavy elements to lighter elements,
such as Fe/C, Fe/O, and Mg/C would be too large.  In this model, Ti,
Co, Zn and Sc are still underproduced. However, these elements may be
enhanced efficiently in the aspherical explosions (Maeda et al. 2002;
Maeda \& Nomoto 2003ab).

\subsection{EMP Stars with a Typical Abundance Pattern}

\begin{figure}[t]
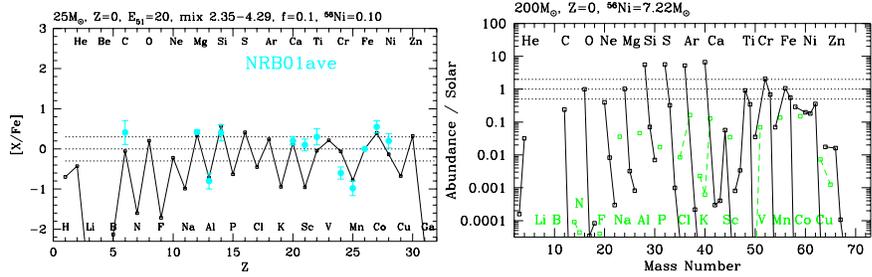

  \begin{center}
  \begin{minipage}[t]{0.49\textwidth}
		\includegraphics[width=1.\textwidth]{nrbave.epsi}
  \end{minipage}
  \begin{minipage}[t]{0.49\textwidth}
		\includegraphics[width=1.\textwidth]{pi200z0rat.epsi}
  \end{minipage}  \end{center}
\caption{
(left) Averaged elemental abundances of stars with [Fe/H] $= -3.7$
(Norris et al. 2001) compared with a theoretical supernova yield
(Umeda \& Nomoto 2003b). (right) Yields of a pair-instability
supernova from the 200 \ms\ star (Umeda \& Nomoto 2002a).
\label{figpi}}
\end{figure}

Similarly, the "mixing and fall back" process can reproduce the
abundance pattern of the typical EMP stars without enhancement of C
and N.  Figure~\ref{figpi} (left) shows that the averaged abundances
of [Fe/H] $= -3.7$ stars in Norris et al. (Norris et al. 2001) can be
fitted well with the model of 25 \ms\ and $E_{51} =$ 20 but larger
fraction ($\sim$ 10\%) of the processed materials in the ejecta.  This
yield (Umeda \& Nomoto 2003b) is recommendable as averaged
core-collapse SN yields for the use of chemical evolution models.

\subsection{Aspherical Explosions}

\begin{figure}[t]
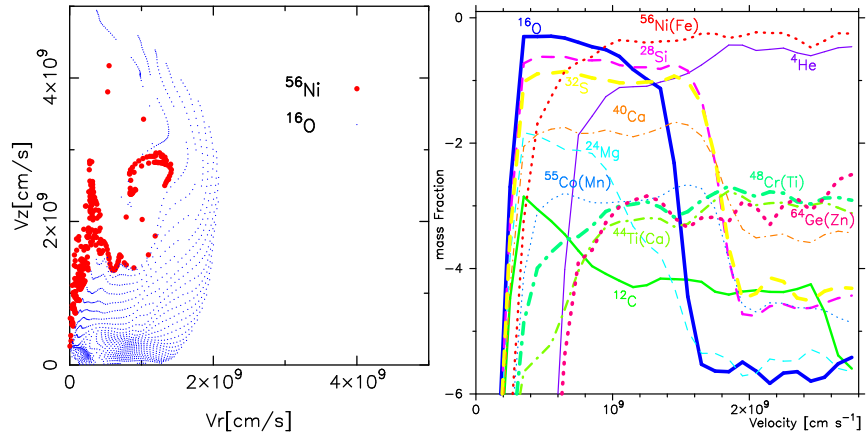

	\begin{center}
		\begin{minipage}[t]{0.49\textwidth}
		\includegraphics[width=1.0\textwidth]{maeda_fig2a.epsi}
		\end{minipage}
		\begin{minipage}[t]{0.49\textwidth}
		\includegraphics[width=1.0\textwidth]{maeda_fig2b.epsi}
		\end{minipage}
	\end{center}
	\caption{Left: Distributions of $^{56}$Ni (which decays into
$^{56}$Fe: filled circles) and $^{16}$O (dots).  The mass elements in
which the mass fraction of each isotope exceeds 0.1 are plotted.
Right: Mass fractions of selected isotopes in the velocity space along
the $z$-axis (Maeda \& Nomoto 2003ab).
\label{fig:maeda}}
\end{figure}

The ``mixing and fall-back'' effect may also be effectively realized
in non-spherical explosions accompanying energetic jets (e.g., Maeda
\& Nomoto 2002, 2003ab).  Compared with the spherical model with the
same $M_{\rm cut}(i)$ and $E$, the shock is stronger (weaker) and thus
temperatures are higher (lower) in the jet (equatorial) direction.  As
a result, a larger amount of complete Si-burning products are ejected
in the jet direction, while only incomplete Si-burning products are
ejected in the equatorial direction (Fig. \ref{fig:maeda}).  In total,
complete Si-burning elements can be enhanced (Maeda \& Nomoto 2003ab).

The jet-induced explosion results in angle-dependent distribution of
nucleosynthetic products as shown in Figure \ref{fig:maeda}. The
distribution of $^{56}$Ni (which decays into $^{56}$Fe) is elongated
in the jet direction, while that of $^{16}$O is concentrated in the
central region.

Zn and Co are ejected at higher velocities than Mn and Cr, so that the
latter accrete onto the central remnant more easily. As a consequence,
[Zn/Fe] and [Co/Fe] are enhanced, while [Mn/Fe] and [Cr/Fe] are
suppressed.

\section{The First Stars}

It is of vital importance in current astronomy to identify the first
generation stars in the Universe, i.e., totally metal-free, Pop III
stars.  The impact of the formation of Pop III stars on the evolution
of the Universe depends on their typical masses.  Recent numerical
models have shown that, the first stars are as massive as $\sim$ 100
\ms\ (Abel, Bryan, \& Norman 2002).  The formation of long-lived low 
mass Pop III stars may be inefficient because of slow cooling of metal
free gas cloud, which is consistent with the failure of attempts to
find Pop III stars.

If the most Fe deficient star, HE0107-5240, is a Pop III low mass star
that has gained its metal from a companion star or interstellar matter
(Yoshii 1981), would it mean that the above theoretical arguments
are incorrect and that such low mass Pop III stars have not been
discovered only because of the difficulty in the observations?

Based on the results in the earlier section, we propose that the first
generation supernovae were the explosion of $\sim$ 20-130 \ms\ stars and
some of them produced C-rich, Fe-poor ejecta.  Then the low mass star
with even [Fe/H] $< -5$ can form from the gases of mixture of such a
supernova ejecta and the (almost) metal-free interstellar matter,
because the gases can be efficiently cooled by enhanced C and O ([C/H]
$\sim -1$).

\begin{table}[t]
\caption{The results of the stability analysis for Pop III and Pop I
stars.  $\bigcirc$ and $\times$ represent that the star is stable and
unstable, respectively.  The $e$-folding time for the fundamental mode
is shown after $\times$ in units of $10^4$yr (Nomoto et al. 2002a).}
\begin{center}
\footnotesize
\begin{tabular}{ccccccc}
\hline \hline
{\large mass ($M_\odot$)} &{\large 80}&{\large 100}&{\large 120}&{\large 150} &{\large 180} &{\large 300} \\ \hline
{\large Pop III} &{\large $\bigcirc$ }&{\large $\bigcirc$ }&{\large $\bigcirc$ }&{\large $\times$ (9.03)} &{\large

 $\times$ (4.83)} &{\large $\times$ (2.15)} \\ 
{\large Pop I }&{\large $\bigcirc$ }&{\large $\times$ (7.02)} &{\large $\times$ (2.35)} &{\large $\times$ (1.43)} &{\large
 $\times$ (1.21)} &{\large $\times$ (1.71)} \\ \hline 
\end{tabular}
\end{center}
\end{table}

\subsection {Pair-Instability Supernovae}

We have shown that the ejecta of core-collapse supernova explosions of
20-130 \ms\ stars can well account for the abundance pattern of EMP
stars. In contrast, the observed abundance patterns cannot be
explained by the explosions of more massive, 130 - 300 \ms\
stars. These stars undergo pair-instability supernovae (PISNe) and are
disrupted completely (e.g., Umeda \& Nomoto 2002a; Heger \& Woosley
2002), which cannot be consistent with the large C/Fe observed in
HE0107-5240 and other C-rich EMP stars.  The abundance ratios of
iron-peak elements ([Zn/Fe] $< -0.8$ and [Co/Fe] $< -0.2$) in the PISN
ejecta (Fig.~\ref{figpi}; Umeda \& Nomoto 2002a; Heger \& Woosley
2002) cannot explain the large Zn/Fe and Co/Fe in the typical EMP
stars (McWilliam et al. 1995; Primas et al. 2000; Norris et al. 2001)
and CS22949-037 either.  Therefore the supernova progenitors that are
responsible for the formation of EMP stars are most likely in the
range of $M \sim 20 - 130$ \ms, but not more massive than 130 \ms.
This upper limit depends on the stability of massive stars as will be
discussed below.

\subsection{Stability and Mass Loss of Massive Pop III Stars}

To determine the upper limit mass of the Zero Age Main Sequence
(ZAMS), we analyze a linear non-adiabatic stability of massive
($80M_{\odot}$ - $300M_{\odot}$) Pop III stars using a radial
pulsation code (Nomoto et al. 2002a).  Because CNO elements are absent
during the early stage of their evolution, the CNO cycle does not
operate and the star contracts until temperature rises sufficiently
high for the $3\alpha$ reaction to produce $^{12}$C.  We calculate
that these stars have $X_{\rm CNO} \sim 1.6 - 4.0\times10^{-10}$, and
the central temperature $T_{\rm c}\sim1.4\times10^8$K on their ZAMS.
We also examine the models of Pop I stars for comparison.

Table 1 shows the results for our analysis. The critical mass of ZAMS
Pop III star is $128M_{\odot}$ while that of Pop I star is
$94M_{\odot}$.  This difference comes from very compact structures
(with high $T_{\rm c}$) of Pop III stars.

Stars more massive than the critical mass will undergo pulsation and
mass loss. We note that the $e$-folding time of instability is much
longer for Pop III stars than Pop I stars with the same mass, and thus
the mass loss rate is much lower. These results are consistent with
Ibrahim, Boury, \& Noels (1981) and Baraffe, Heger, \& Woosley (2001). 
However, the absence of the indication of PISNe may imply that these
massive stars above 130 \ms\ undergo significant mass loss, thus
evolving into Fe core-collapse rather than PISNe.

\section{Discussion}

We have first shown that signatures of hypernova nucleosynthesis are
seen in the abundance patterns in extremely metal-poor (EMP) stars.
We suggest that hypernovae of massive stars may make important
contributions to the Galactic (and cosmic) chemical evolution,
especially in the early low metallicity phase.  The IMF of Pop III
stars might be different from that of Pop I and II stars, and that
more massive stars are abundant for Pop III.

We have also shown that the most iron-poor star as well as other
C-rich EMP stars is likely to be enriched by massive supernovae that
are characterized by relatively small Fe ejection. Such supernovae are
not just hypothetical but have been actually observed, forming a
distinct class of type II supernovae ("faint supernovae": Nomoto et
al. 2003ab). The proto-type is SN1997D, which is very under luminous
and shows quite narrow spectral lines (Turatto et al. 1998) (also
SN1999br; Zampieri et al. 2003): These features are well modeled as an
explosion of the 25 \ms\ star with small explosion energy $E_{51} =$
0.4.  On the other hand, typical EMP stars without enhancement of C
and N correspond to the abundance pattern of energetic supernovae
("Hypernovae": Nomoto et al. 2003ab).

For both cases black holes more massive than $\sim$ 3 - 10 \ms\ must be
left as a result of fallback, suggesting that the copious formation of
the first black holes from the first stars. These black holes may
consist of some of the dark mass in the Galactic halo. In our scenario,
HE0107-5240 with [Fe/H] $= -5.3$ was formed from C, O-enhanced gases with
[C,O/H] $\sim -1$. With such enhanced C and O, the cooling efficiency is
large enough to form small mass stars. As far as the low mass EMP stars
are C-rich, therefore, their small masses are consistent with the
massive Pop III star formation. Rather their C-richness implies that Pop
III stars that are responsible for their formation are massive enough to
form (the first) black holes.

\section*{Acknowledgements}
Detailed yields are seen at 
http://supernova.astron.s.u-tokyo.ac.jp/~umeda/data.html.  
This work has been supported in part by the grant-in-Aid for COE
Scientific Research (14047206, 14540223) of the Ministry of Education,
Science, Culture, Sports, and Technology in Japan.

\end{document}